%% file: vovk.tex
\documentclass[iop]{emulateapj}
\usepackage{longtable}
\usepackage[utf8x]{inputenc}
\usepackage{graphicx}
\usepackage{color}

\def\gr{$\gamma$-ray}

\begin{document}

\title{Variability of gamma-ray emission from blazars on the black hole timescales}

\author{Ie.~Vovk and A.~Neronov}
\affil{ISDC Data Centre for Astrophysics, Ch. d'Ecogia 16, 1290, Versoix, Switzerland}

\begin{abstract}
We investigate the variability properties of blazars in the GeV band using the data of the Fermi/LAT telescope. We find that blazars exhibit variability on the scales down to the minimal timescale resolvable by Fermi,  which is a function of the peak photon count rate in the LAT. This implies that the real minimal variability timescales for the majority of blazars are typically shorter than those resolvable by the LAT. We find that for several blazars these minimal variability timescales reach those associated to the blazar central engine, the supermassive black hole. At the same time, none of the blazars exhibits variability on the timescale shorter than the  black hole horizon light crossing time and/or the period of rotation around the last stable circular orbit. Based on this fact, we argue that the timing properties of the \gr\ signal could be determined by the processes in the direct vicinity of the supermassive black hole.
\end{abstract}


\section{Introduction}

Blazars are the subclass of radio-loud AGNs, for which the jet is closely aligned with the line of sight. They emit most of their power in the X- to $\gamma$-ray domain. This high-energy emission most probably originates in the jets and is produced via the Inverse Compton scattering of low-energy photons. 

Jets of the radio-loud AGNs are known to span over many orders of magnitude in distance, from the Astronomical Unit scale of the AGN central engine, the supermassive black hole (SMBH), up to the parsec to megaparsec scales. The exact location of the "blazar emission zone" in the jet is not known.  If  the \gr\ flux originates  from within the Broad Line Region (BLR), where the radiation fields are intense, interactions of the \gr s with low energy background photons leads to the $e^\pm$-pair production thus preventing the $\gamma$-rays from escaping the source \citep{BLR_absorption}. This problem could be avoided if the site of production of the $\gamma$-rays is located outside from the BLR, parsecs away from the central SMBH. 

However, the possibility of location of the emission region at large distances poses a problem for the explanation of the  observed variability timescales.  Over the last few years, a number of blazars were found to exhibit variability on very short timescales, down to several minutes \citep{Aharonian_PKS_2155, Mrk501, PKS_2155_HESS,Foschini_fast_var, NV_OJ287, Sbarrato_var_study}. Simple causality arguments indicate that the short variability timescales could be best explained by the compactness of the \gr\ emission regions, which appears natural if the blazar emission zone is situated close to the central engine. In fact, \gr\ emission produced at the distances $D\sim \Gamma^2R_{\rm CE}$, where $\Gamma\sim 3-30$ is the bulk Lorentz factor of the jet and $R_{\rm CE}$ is the size of the blazar central engine, could be variable at the timescale as short as the light crossing time of the central engine, $t_{\rm CE}=R_{CE}/c\sim 10^3\left[R_{\rm CE}/2\mbox{ AU}\right]\mbox{ s}$ \citep{Celotti_1998}.

Although fast variability provides a strong argument in favor of the $\gamma$-ray production within the BLR regions, close to the central engine, alternative possibilities, usually invoking very compact emission regions embedded in the larger scale jet have been considered to explain the fast variability phenomenon \citep{Ghisellini_needles-in-jet, Ghisellini_streams, Giannios_jet-in-jet, Piran_jet-in-jet}.

Accumulation of a larger sample of fast variable blazars is needed to distinguish between the two possibilities. If the fast variability is due to the location of the \gr\ emission region close to the blazar central engine, one generically expects that the shortest variability timescale is determined by the central engine's size. In this case the minimal variability timescales of blazars are expected to scale proportionally to the mass of the central SMBH. Otherwise, if the shortest timescales are related to the compact emission regions at large distances, formed as a result of the development of instabilities intrinsic to the jet, the minimal variability timescales are not expected to depend on the size of the SMBHs powering the jets. 

In this work we focus on the investigation of the minimal variability timescales of a large number of blazars observed by the Large Area Telescope (LAT) on board of Fermi satellite \citep{atwood09}. LAT is operating in the all-sky survey mode and is well suited for the monitoring of activity of blazars at different timescales in the 0.1-100~GeV energy band. At the same time, the relatively small effective area of LAT limits the photon statistics at the shortest minute-hour-day timescales expected if the fastest variability scale is determined by the SMBH. Below we study the variability properties of a large number of blazars which are detected by LAT in a two-year exposure \citep{2FGL} and for which estimates of the mass of the SMBH exist. We demonstrate that although the minimal detected variability timescales are generically limited by the LAT sensitivity, for some $\sim 10$ brightest sources LAT sensitivity is sufficient to probe the fundamental SMBH timescale. For these brightest sources, variability at the timescale comparable to the period of rotation around the last stable orbit is in fact observed. For 3 sources among these, the detected minimal variability timescales are found to be truly minimal, intrinsic to the sources. At the same time, we find that LAT sensitivity is currently not sufficient to confirm or rule out the correlation of the minimal variability timescales with the SMBH mass. We argue that accumulation of larger exposure with LAT will increase the statistics of the bright blazar flares during which the sources exhibit the short-time-scale variability. This should open the possibility for testing the minimal variability timescale -- black hole mass correlation. 

\section{The blazar sample}

The Fermi Large Area Telescope (Fermi/LAT) is a space borne detector, sensitive to the $\gamma$ rays approximately from 100 MeV to 300 GeV. This instrument operates in all-sky scanning mode, completing the survey in two revolutions ($\sim3.2$ hours). The Fermi/LAT two-year source catalogue \citep{2FGL}  contains $\sim 1800$ sources, among which $\sim 1000$ are associated with AGNs.

For the purpose of this work, we select blazars from the LAT two-year catalogue for which the estimates of the black hole masses exist. The mass estimates mainly come from observations in the optical domain. Several alternative methods for the mass estimates are used. One possibility is to use the observed correlation between the SMBH mass and velocity dispersion in the central stellar bulge observed in the nearby galaxies. Measuring the bulge velocity dispersion through the spectroscopic observations of the blazar host galaxy one could deduce an estimate of the SMBH mass \citep{Sigma-mass_relation}.  Otherwise,  the black hole mass could be estimated from a combination of measurement of the width of the atomic line emission from the BLR and the time delays between the continuum and line emission (reverberation mapping technique) \citep{Reverberation_mapping}. 

%%%%%%%%%%%%%%%%%%%%%%%%%%%%%%%%%%%%%%%%%%%%%%%%%%%%
% === Table with the source list ===
\input{Table_of_sources.tex}
% \input{table.tex}
%%%%%%%%%%%%%%%%%%%%%%%%%%%%%%%%%%%%%%%%%%%%%%%%

In the case of blazars, which form the majority of the AGNs observed in the GeV domain, it is often the case that the velocity dispersion of the bulge can not be measured due to the strong contamination with the emission of the AGN itself. In this case less direct estimates are used to infer the mass of the SMBH. These estimates are based on the so-called \textquotedblleft Fundamental plane\textquotedblright, that gives relation between the effective radius, average surface brightness and the velocity dispersion of the early-type galaxies \citep{FundPlane}. A similar correlation was found between the optical R-band luminosity of the bulge and the mass of the central SMBH \citep{McLure}.

For the purpose of this work we combined from the literature \citep{FalomoCat, FundPlane, Sephane_3C273, XieCat, BLLac170} the list of 195 AGN with the SMBH masses measured using different methods. We further selected those sources, that were observed by the Fermi/LAT telescope. Among those, only the sources without bright companions within $2^\circ$-radius around their positions were chosen for the analysis. Such selection criterion assures that the variability properties of the source of interest are not affected by the variability of nearby sources (the Point-Spread-Funciton of LAT is several degrees wide in the energy range below $\sim 1$~GeV \footnote{http://www.slac.stanford.edu/exp/glast/groups/\newline /canda/lat\_Performance.htm}). The resulting list of the 86 AGNs, used in present work, is given in table \ref{SrcTable}.

\section{Data analysis}

\subsection{Light curves}

For the analysis we used the \textit{Fermi Science Tools v9r23p1}\footnote{http://fermi.gsfc.nasa.gov/ssc/data/analysis/software/} - a standard software package, provided by the Fermi collaboration to reduce the data, obtained by the Fermi/LAT. We also made use of the Pass 7 version of the photon files, which is stated by the Fermi collaboration as the best dataset for the Fermi/LAT analysis\footnote{http://fermi.gsfc.nasa.gov/ssc/data/analysis/documentation/ \newline /Pass7\_usage.html}. We selected the class 2 photons and applied the cut on the zenith angle of $105^\circ$.

To produce the light curves we used the \textit{gtbin} tool, and computed exposures, corresponding to each bin, with the help of \textit{gtexposure}. We fixed the spectral indices of the sources at the values, presented in the Fermi/LAT two-year source catalogue \citep{2FGL}. The energy range in our analyses was set to 100 MeV - 300 GeV. As we were not interested in the absolute normalization of flux, but were rather kin to detect variability, we set the software to collect photons within a small, $1^\circ$ circle. This is much smaller than the 68\% photon containment radius for 100 MeV, and is roughly equal to the same containment radius at 1 GeV. This way the analysis uses only the central, core part of the PSF, omitting its wings, thus minimizing contribution from the nearby sources.
The light curves we built with the signal-to-noise ratio (SNR) binning, with the value of SNR in each bin fixed at 4. Reducing the noise when the flux is low, such a binning allows to resolve finest details in the light curve during the flares and, thus, to probe the fastest variability of the source.

The above-mentioned procedure is equivalent to the usual aperture photometry with two exceptions. First is that we do not collect all the photons from the source, as we select only the narrow central core of the PSF for our analysis. The second difference comes from the fact, that we do not perform the background subtraction. As the background is not expected to be variable, this should not introduce any additional time scales into our results, but might limit the sensitivity to the fast variability of weak sources.

\subsection{Variability search}

To search for the variability we utilized the structure function approach \citep{SF_intro}. The structure function (SF) of the light curve $x(t)$ with the time lag $\tau$ is defined as:
\begin{equation}
  SF_x(\tau) = \langle (x(t+\tau)-x(t))^2 \rangle
  \label{Eq::SF}
\end{equation}
and represents the amount of variability at a given timescale $\tau$. It was first introduced in astronomy by \cite{SF_intro}, and afterwards successfully used to measure the variability of AGNs (e.g. \cite{SF_Stephane}). The advantage of the SF is that it works in the time domain, making it less sensitive to windowing problems than the Fourier analysis.

There are a few important properties of the structure function, that are crucial in the frames of this paper. First of all, for very small values of $\tau$, far below the shortest variability timescale, the SF is dominated by the noise in the signal and, as the noise is independent from any timescale, the structure function here is constant. For the larger $\tau$, but still below the minimal variability timescale, the SF represents the small-scale linear trends in the signal and grows as $\tau^2$. For the timescales, where the variability occurs, the change of the SF is defined by the type of the variability itself (for example, if a periodicity is present in the signal, the SF will show a dip at the corresponding period). Finally, for the largest values of $\tau$, larger than any variability timescale present in the signal, the structure function forms a plateau at the value, corresponding to the variance of the incoming signal. Thus the overall look of the structure function for a particular light curve is defined by the combination of the minimal and maximal variability timescales, the level of noise and the particular binning, used to produce the light curve. A deeper discussion of the SF properties and caveats was given by \cite{SF_caveats}.

The structure function also provides a direct way of measuring the amplitude of the variability at a given timescale. Ratio of values of SF at two different timescales is the square of the ratio of the corresponding amplitudes (see Eq.~\ref{Eq::SF}). Probably the most convenient way of quantifying the variability amplitude is to compare the corresponding value of SF with its plateau value in the limit of large $\tau$, which represents the overall variance of the light curve.

To find the variability one has to compare the observed inclination slope of the logarithmically scaled SF ($\alpha$=$d\log(SF)/d\log(\tau)$) with 2 (linear grows) and 0 (below the minimal or above the maximal variability timescale). The linear growth, which not considered as variability here, may occur because of two different reasons. First, expansion of almost any signal into the Taylor series leads to the dominant linear term for sufficiently small values of $\tau$, so generally one expects the slope of SF to be 2 for these values of $\tau$ for any light curve. If, one the other hand, the light curve itself is a simple linear increase of flux, then there is no timescale to be associated with this growth, and the minimal variability timescale can not be derived. The apparent slope of the SF is affected by the noise, introduced due to the limited number of photons in each bin of the light curve. Before checking the slope of the SF one has to remove the contribution from this noise. In the this work we assumed that the noise is purely Poissonian, corresponding to the amount of counts in each time bin. To calculate its influence on the resulting SF, we produced a number of simulated light curves, for which the time bins were taken as in the initial one, and number of photons in each bin was taken from the Poisson distribution with the mean equal to the initial number of photons in this bin. This way we expected to add to the light curve the very same noise as the one already present there, and, as a result, to double the value of noise plateau of the SF. The resulting difference between the averaged over the simulations SF and the initial one should be equal to value, added to the SF by the noise. Having subtracted this value from the initial SF, we obtained the noise-subtracted structure functions for each source. At the same time, the distribution of the values of simulated SFs for every given $\tau$ defined the uncertainties of the structure function, derived from the original light curve.

%%%%%%%%%%%%%%%%%%%%%%%%%%%%%%%
\begin{figure}
 \begin{center}
    \includegraphics[width=0.49\linewidth]{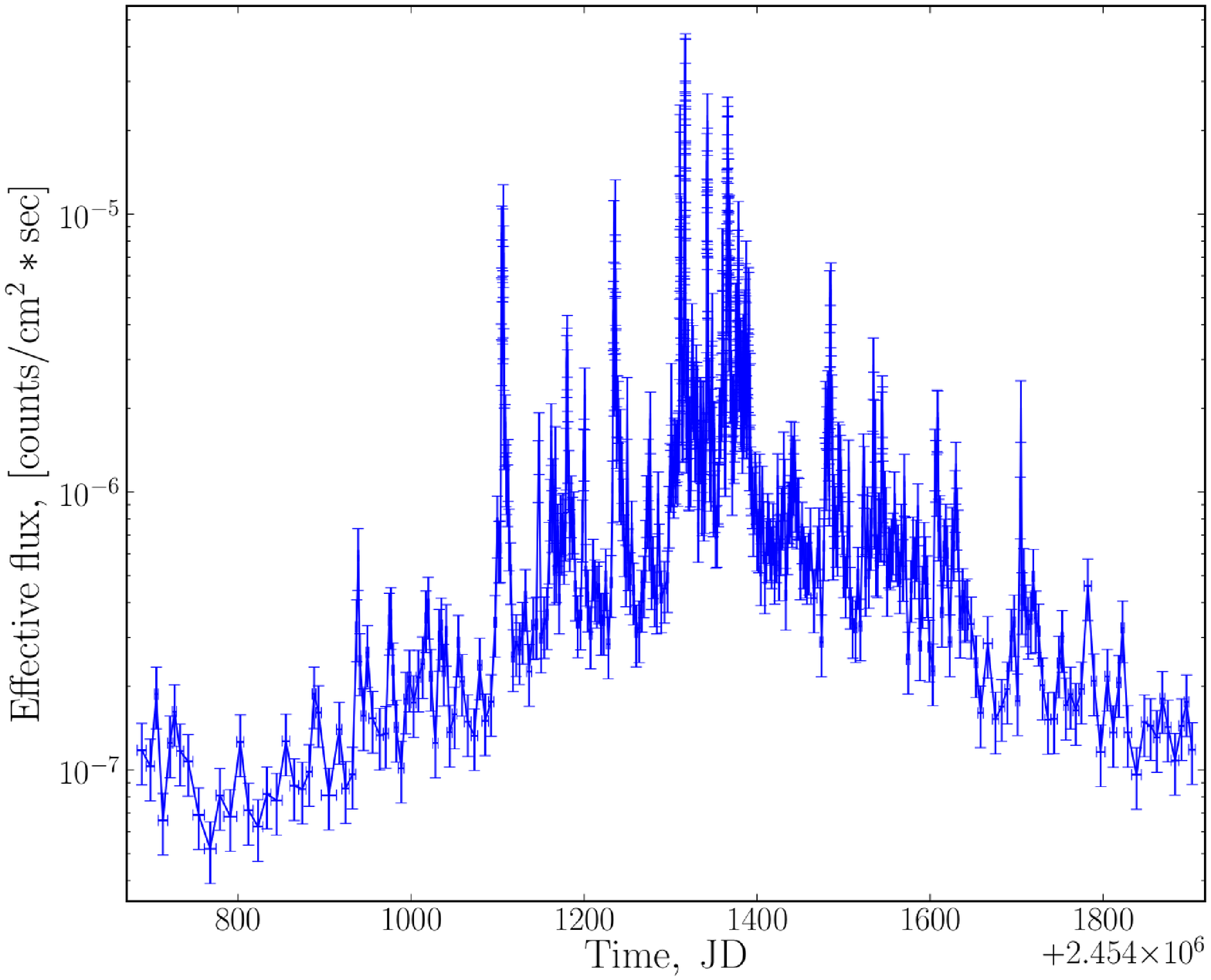}
    \includegraphics[width=0.49\linewidth]{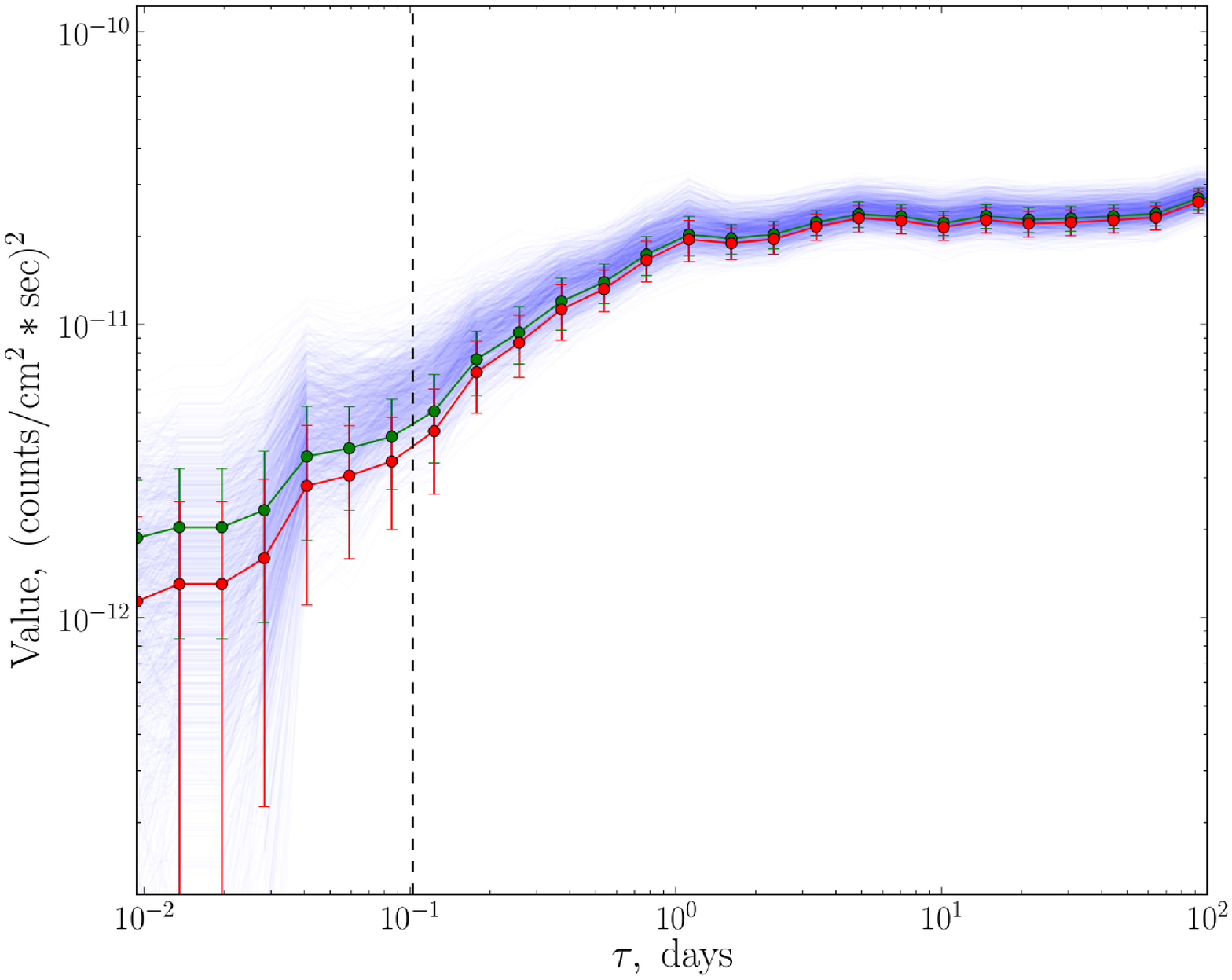}
    \includegraphics[width=0.49\linewidth]{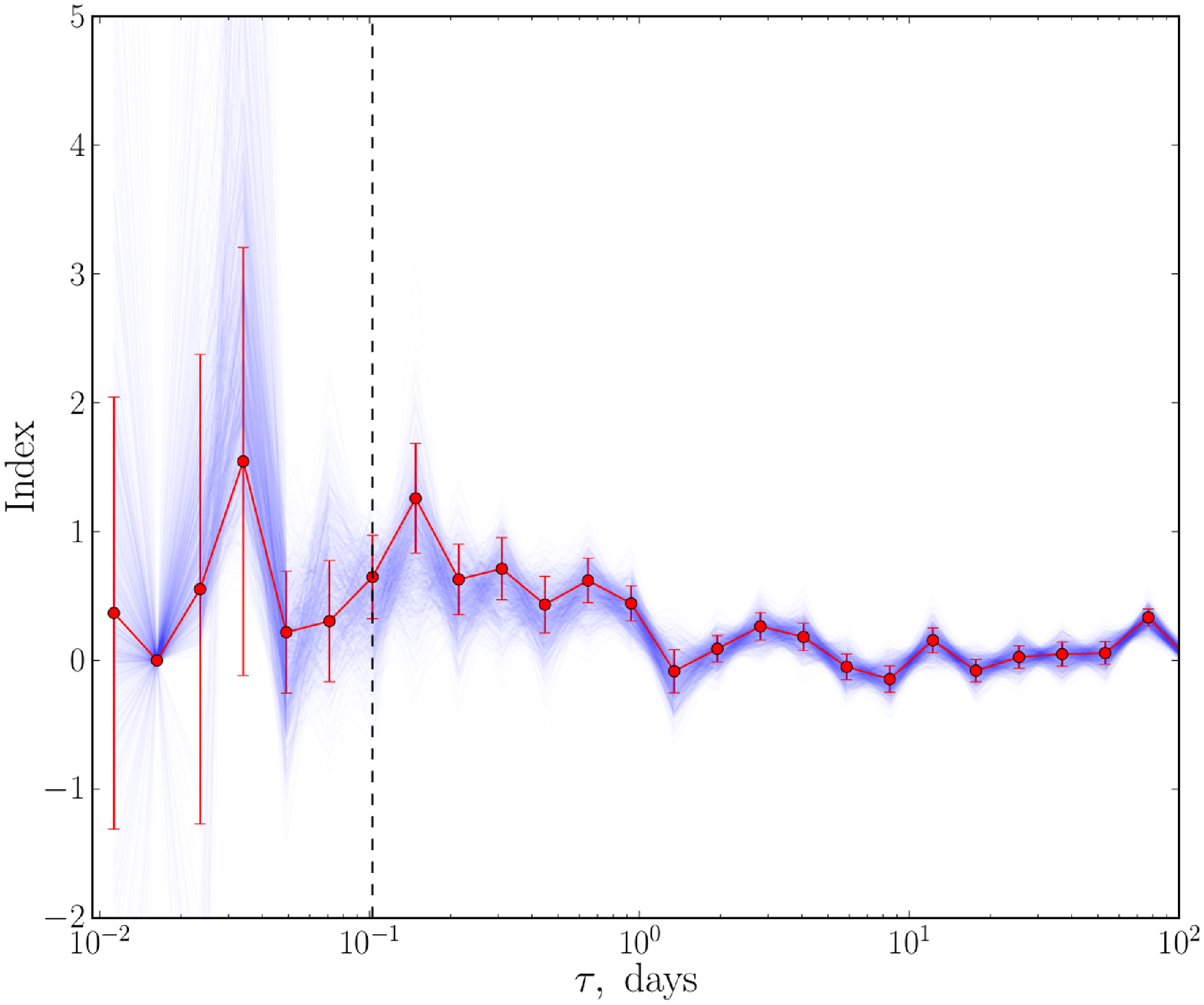}
 \end{center}
 \caption{Example of the complete analysis cycle -- 4C +21.35. Top left panel: the SNR-binned light curve. Top right panel: structure function before (green) and after (red) the noise removal. Bottom panel: structure function logarithmic inclination index. Light blue lines represent the distribution of simulated SFs (central panel) and their indices (bottom panel), used to compute the error bars (95\% confidence range). Minimal present variability timescale is 0.1 days.}
 \label{ExampleSF}
\end{figure}
%%%%%%%%%%%%%%%%%%%%%%%%%%%%%%%

Uncertainties of the logarithmic inclination indices $\alpha$ of the structure function were also derived from these simulations. Because the computed error bars of SF are correlated (as every simulation resembles the original light curve), we computed the inclination indices for each simulated SF and then used their distribution for each value of $\tau$ to compute the 95\% confidence error bars for the indices of the initial structure function.

We then compared all the indices of the SF with 2 and 0, as it was described above. The value of $\tau$ of the first bin, that, with the account of the error bars, would have an index different from these two reference values, was considered to be an estimate of the minimal variability timescale, present in the light curve.

As an example, the light curve, structure function and structure function indices of 4C +21.35, obtained in this way, are shown in Fig.~\ref{ExampleSF}.

The minimal variability timescale, derived in this manner, may depend on the particular way the bins were chosen. Binning with a constant SNR essentially collects the photons in groups of $N=SNR^2$, which is 16 in our case (SNR=4). Thus, in principle, we have a freedom of choosing these bins in 16 different ways. To account for this, for each source we produced 16 different light curves, each starting from the second photon of the previous one, in this way going over all the possible combinations. The timescales, derived for these light curves, then gave us the uncertainty to which the minimal variability is determined. If for any of those light curves variability was not found, we assumed that the source was not variable over the time interval covered by our analysis.

\section{Results}

\begin{figure}
  \center \includegraphics[width=\linewidth]{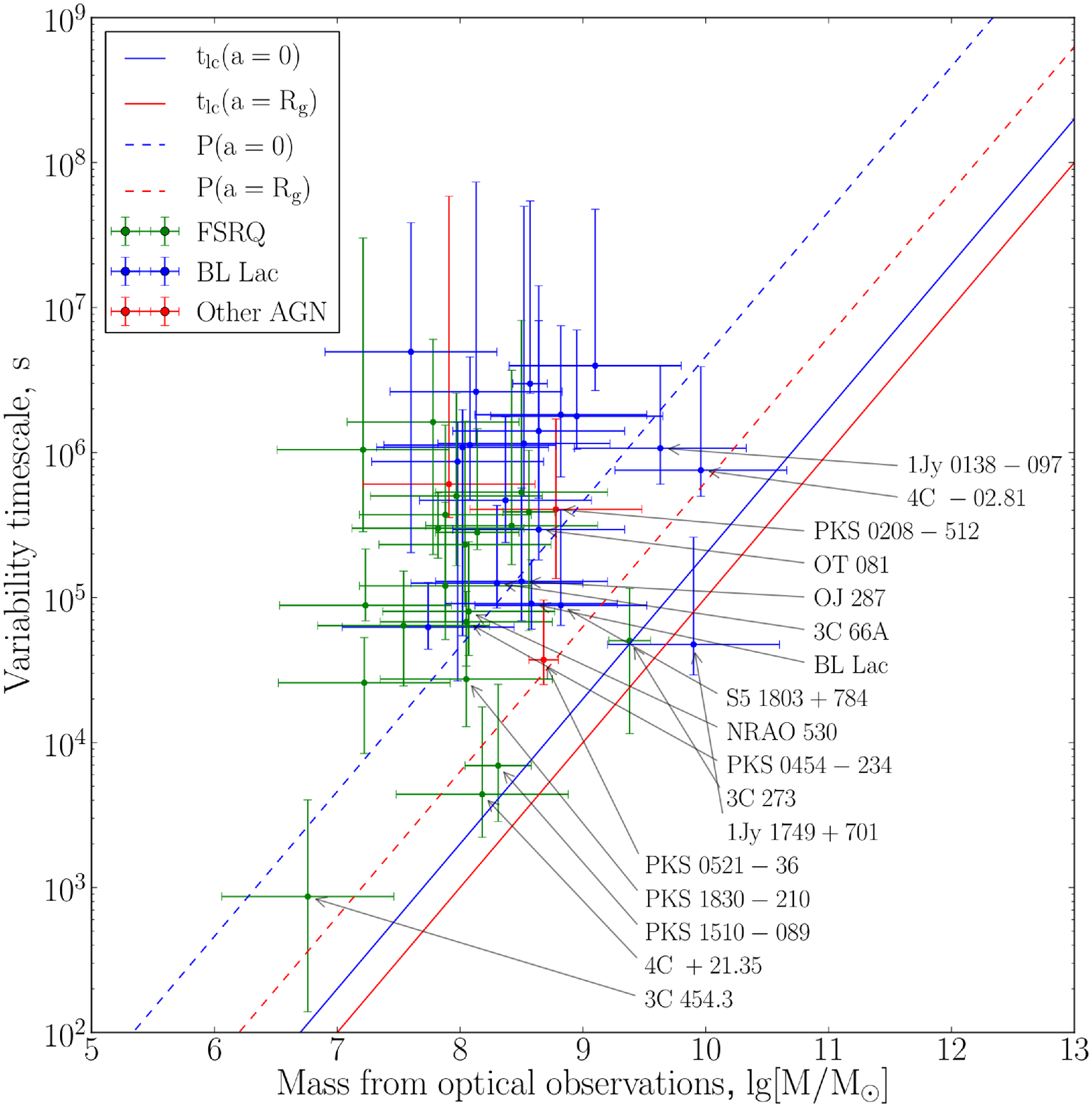}
  \caption{Measured minimal variability timescales versus the mass of the central black hole for the AGNs from table \ref{SrcTable}. Solid lines represent the light-crossing times for maximally rotating (red) and non-rotating (blue) black holes, dashed lines -- the same for the period of the last stable prograde orbit.}
  \label{Results_types}
\end{figure}

The derived minimal variability timescales, plotted versus the corresponding SMBH mass, are shown in Fig.~\ref{Results_types}. As it can be seen from this figure, the majority of blazars in our sample vary on timescales significantly larger than the light-crossing times of their SMBHs (shown by the solid straight lines)
\begin{eqnarray}
  \label{eq_t_lc}
  t_{lc}&=& 2(R_g+\sqrt{R_g^2-a^2})/c \\
 & \simeq &\left\{ \begin{array}{lr}
                                                      10^3 \left(\frac{\displaystyle M_{BH}}{\displaystyle 10^8 M_\odot}\right) s,  &  a=R_g \\
                                                      2\times 10^3 \left(\frac{\displaystyle M_{BH}}{\displaystyle 10^8 M_\odot}\right) s,  &  a=0   \\
                                                    \end{array}                                           
                                            \right.\nonumber
\end{eqnarray}
where $R_g = GM_{BH}/c^2$ and $a = J_{BH}/M_{BH}c^2$ is defined by the angular momentum $J_{BH}$ of the black hole and lies in the range $0 \leq a \leq R_g$.

 A small fraction (some $\sim 10\%$) of the sources demonstrate variability timescales as small the period of the last stable orbit (shown by the dashed lines in Fig. \ref{Results_types})
\begin{equation}
  \label{eq_period}
  P(r_{min}) = \left\{ \begin{array}{lr}
                        \frac{\displaystyle4\pi R_g}{\displaystyle c} \simeq 6\times 10^3 \left(\frac{\displaystyle M_{BH}}{\displaystyle10^8 M_\odot}\right) s,            & a = R_g \\
                        \frac{\displaystyle 12 \sqrt{6}\pi R_g}{\displaystyle c} \simeq 5\times 10^4 \left(\frac{\displaystyle M_{BH}}{\displaystyle 10^8 M_\odot}\right) s,  &a = 0   \\
                       \end{array} \\
               \right.
\end{equation}

From Fig. \ref{Results_types} one can notice that different blazar types, BL Lacs and Flat Spectrum Radio Quasars (FSRQ), exhibit variabilities at different timescales, with FSRQs being systematically variable at shorter timescales. It is not clear a-priori, if this reflects the difference in the mechanisms of \gr\ emission in these types of objects or is related to a selection bias or an instrumental effect in LAT. 

%%%%%%%%%%%%%%%%%%%%%%%%%%%%%%%%%%%%%%%%%%%%%%%%%%
%%%%%%%%%%%%%%%%%%%%%%%%%%%%%%%%%%%%%%%%%%%%%%%%%%

It is also not clear a-priori if the sources which do not reveal variability on the timescales as short as the black hole light crossing time and/or the period of rotation around the last stable orbit are just not variable on these timescales or the sensitivity of the LAT is not sufficient for the detection of variability on such short timescales. 

\begin{figure}
  \center \includegraphics[width=\linewidth]{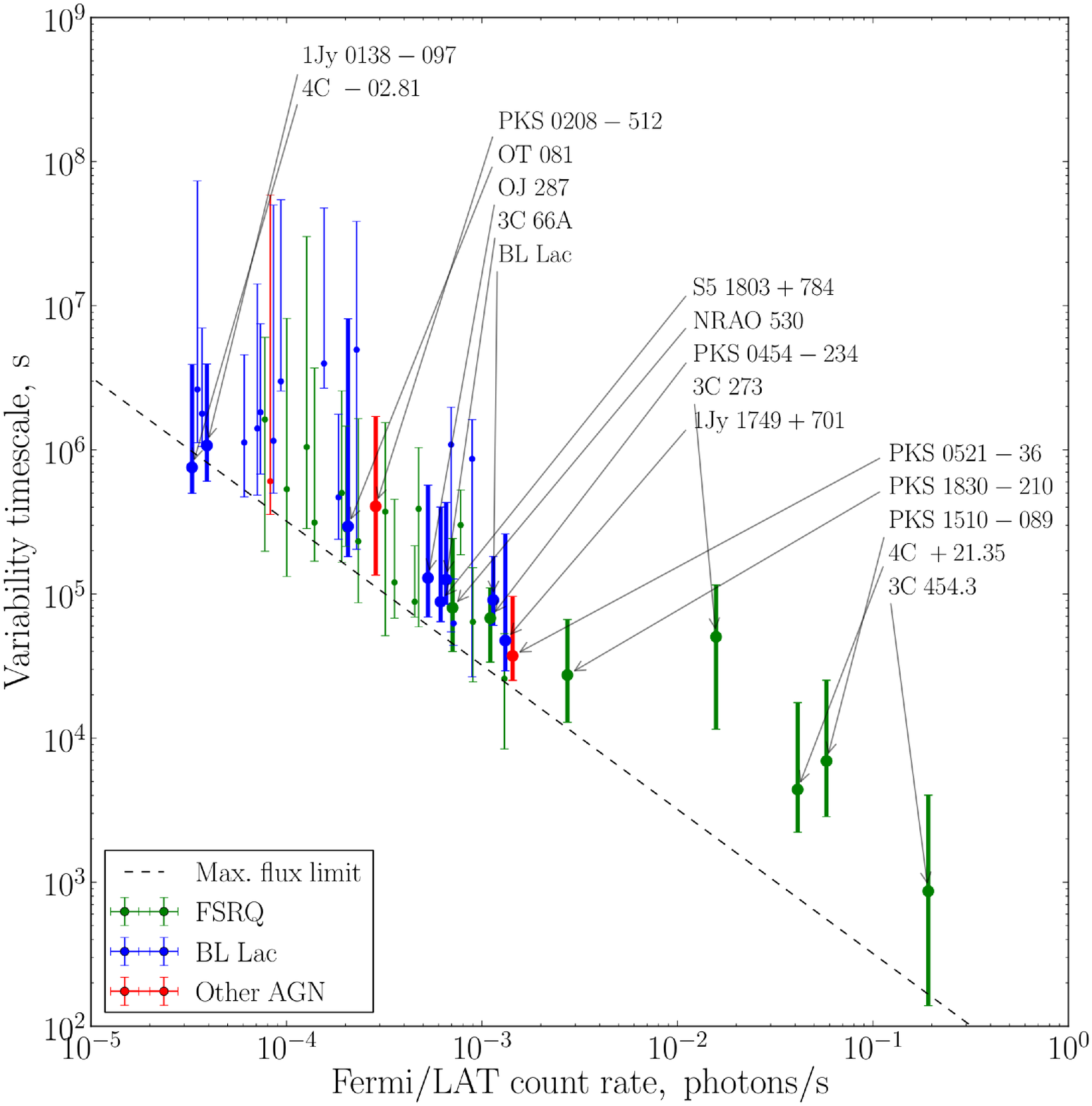}
  \caption{Measured minimal variability timescales versus the maximal count rate found in present analysis. Dashed line represents the minimal variability timescale measurable with a given count rate. Sources found to have variability on the timescales comparable to/shorter than the period of the last stable orbit are marked with bold.}
  \label{Results_countrates}
\end{figure}

In order to investigate the sensitivity limit of LAT, we plot in Fig.~\ref{Results_countrates} the dependence of the minimal variability timescale on the peak count rate in the LAT.  From this figure one could see a trend that the minimal measured variability timescale decreases with the increasing peak count rate. This could be readily explained by the fact that detection of variability at a given timescale requires sufficient signal statistics. The minimal statistics in our case is 16 photons in one time bin. The minimal detectable variability time could not be shorter than the time necessary for collection of 32 photons (there should be at least two time bins to compare the fluxes). This minimal detectable scale is shown by the dashed line in Fig.   \ref{Results_countrates}. One could see that the detected variability timescales follow the minimal detectable timescale line. Thus, the non-detection of variability of the signal from most of the blazars on the SMBH timescales is not related to the absence of such variability, but is rather due to the sensitivity limit of LAT. 

From the same figure one could see that the difference in the variability timescales of FSRQs and BL Lacs is due to the fact that FSRQs are on average brighter than BL Lacs in the LAT energy band. Higher count rates of FSRQs allow the detection of shorter variability timescales for these sources.

One can also notice from Fig.~\ref{Results_countrates}, that for 3 blazars, namely 3C~273, 4C +21.35 and PKS~1510-089, the sensitivity of Fermi/LAT allows the detection of an order of magnitude faster variability, than really observed. This suggests that the found minimal variability timescales of these sources correspond to the truly minimal timescales, defined by the physics of their emission regions. Analysis results for these sources are briefly summarized below.

%%%%%%%%%%%%%%%%%%%%%%%%%%%%%%%%%%%%%%%%%%%%%%%%%%
%%%%%%%%%%%%%%%%%%%%%%%%%%%%%%%%%%%%%%%%%%%%%%%%%%

\begin{figure}
  \center \includegraphics[width=\linewidth]{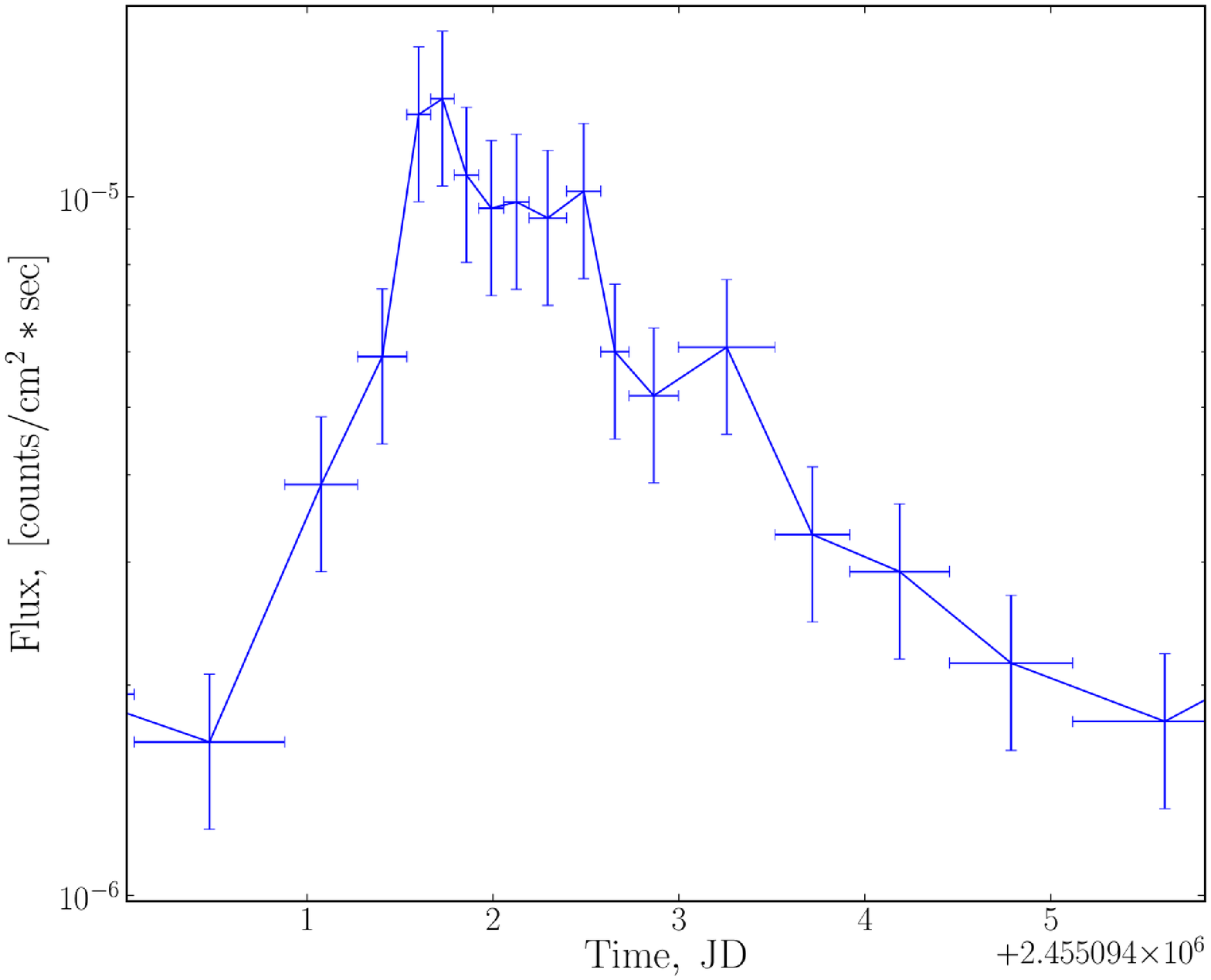}
  \caption{A part of the light curve of 3C 273 around the 21th of September 2009 (JD 2455096), with the bright flare in it.}
  \label{3C_273_flare_zoom}
\end{figure}

\begin{figure}
  \center \includegraphics[width=\linewidth]{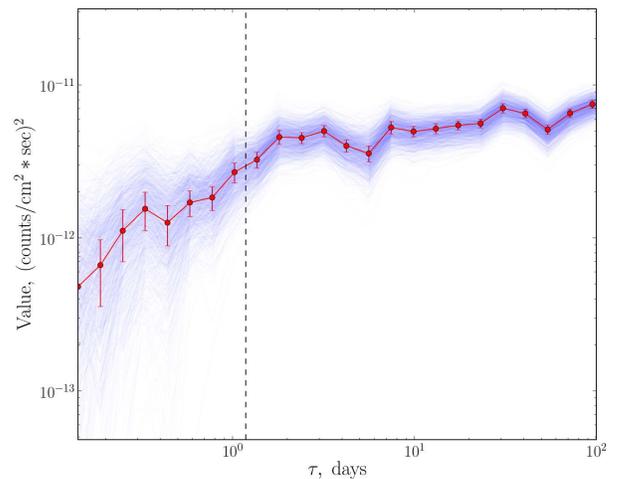}
  \caption{Structure function of the light curve of 3C~273, partially shown in Fig.~\ref{3C_273_periodicity}. Error bars correspond to 68\% confidence ranges. The structure functions from the simulated light curves, used to determine the error bars, are shown with light blue. Minimal detected variability timescale -- $\approx1.18$~days -- is marked with the vertical dashed line. A dip in the structure function is seen at $\tau \sim 5$ days.}
  \label{3C_273_sf}
\end{figure}

\underline{\textit{3C~273}} was variable by factor $\sim 100$ in terms of flux over the period of the Fermi/LAT observations. This highly variable source is also one of the brightest in our sample, thus allowing to study the variability at the smallest timescales. This is demonstrated in Fig.~\ref{3C_273_flare_zoom}, where one of the bright flares, happened on the 21th of September 2009, is shown. As it can be seen from this figure, brightness of this flare ensures that the sufficient amount of photons per bin is achieved even for several hours long intervals, so variability down to a few hours could be, in principle, detectable for this particular light curve, if the variability amplitude is large enough, whereas our analysis indicated, that only $\sim$~day timescales are present (Fig.~\ref{Results_countrates}). This can be seen from Fig.~\ref{3C_273_sf}, which shows the structure function for this particular light curve. Minimal variability timescale, found by our analysis of the entire light curve, is $0.58^{+0.76}_{-0.45}$~days. The corresponding light-crossing time of the central black hole and period of rotation at the last stable orbit can be estimated from equations~\ref{eq_t_lc} and~\ref{eq_period}: $t_{lc} \approx 0.3 \div 0.6$~days and $P(r_{min}) \approx 1.7 \div 14$~days, depending on the black hole rotation momentum.

\begin{figure}
  \center \includegraphics[width=\linewidth]{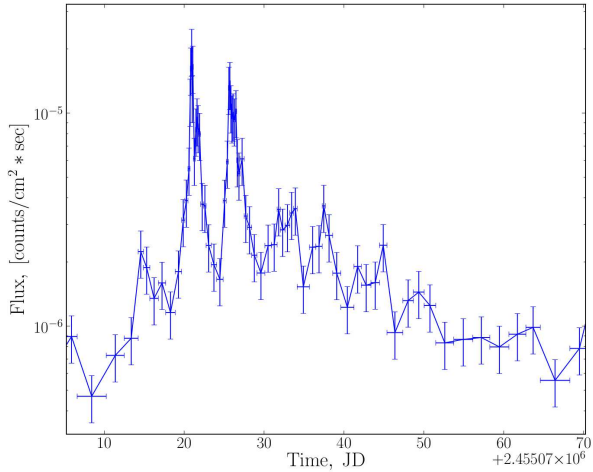}
  \caption{A part of the light curve of 3C 273 in September-November 2009, which shows a possible qausi-periodicity in it.}
  \label{3C_273_periodicity}
\end{figure}

A curious feature, seen in Fig.~\ref{3C_273_sf}, is a dip at the timescale of $\sim 5$~days, which may indicate (quasi)periodic changes in the light curve. Indeed, this behaviour is seen in the light curve, with the most prominent episode shown in Fig.~\ref{3C_273_periodicity}. If being associated with the emission of some blobs of gas, just about to fall onto the black hole, this gives us the period of rotation on the last stable orbit around the central SMBH. As the exact value of this period depends on the black hole's rotation momentum, similar measurements (with a better statistics) could be used to measure the latter.

%%%%%%%%%%%%%%%%%%%%%%%%%%%%%%%%%%%%%%%%%%%%%%%%%%

%%%%%%%%%%%%%%%%%%%%%%%%%%%%%%%%%%%%%%%%%%%%%%%%%%

\underline{\textit{4C~+21.35}} is another highly variable source with bright flares (see Fig.~\ref{ExampleSF}). The minimal variability timescale, determined by our analysis, is $0.050^{+0.022}_{-0.024}$~days or $1.20^{+0.53}_{-0.58}$~hr. Visual inspection of the light curve suggests that it took place during the brightest flare on the 30th of April 2010. As Fermi/LAT operates in the all-sky scanning mode with the scanning period of 3.2~hr, this measurements of the variability timescales below 3~hr may be affected by the fact that telescope did not observe the source constantly. Consequently, it would be more conservative to say that the minimal detected variability timescale lies between $~0.5$~hr, which is the average duration of the observation of a given patch of the sky, and 3.2~hr.

\underline{\textit{PKS~1510-089}} has produced a number of bright flares during the period of the Fermi/LAT operation, with the strongest one taking place on the 19th of October 2011. The minimal variability timescale, derived through the SF analysis, is $0.11^{+0.18}_{-0.06}$~days or $2.6^{+4.4}_{-1.5}$~hr. Similarly to the case of 4C~+21.35, we visually inspected the light curve and concluded that this estimate corresponds to the above mentioned flare. As the derived time scale is smaller than 3.2~hr, this estimate is subject to the same uncertainty, as the estimate for 4C~+21.35. At $\tau \simeq 50$~days structure function exhibits a dip, that continues up to the timescales comparable with the duration of the light curve, which is the largest timescale we can test with our analysis.

%%%%%%%%%%%%%%%%%%%%%%%%%%%%%%%%%%%%%%%%%%%%%%%%%%
\section{Discussion}
%%%%%%%%%%%%%%%%%%%%%%%%%%%%%%%%%%%%%%%%%%%%%%%%%%

Minimal variability timescale provides information on the size and location of the \gr\ emission region and/or on the physical process responsible for the emission. In the case of blazars, the high-energy \gr\ emission might originate from a "blazar emission zone" in the jet at large distance from the central SMBH or, alternatively, from the direct vicinity of the black hole. We have proposed a method to distinguish between these two  scenaria, via a study of correlation of the minimal variability timescale of \gr\ emission with the mass of the SMBH. In the former case no such correlation is expected, because the shortest variability timescales are determined by the size scales and/or Lorentz factors of small inhomogeneities in the blazar jets (possibly formed via development of plasma instabilities or magnetic reconnection events). In the latter scenario, a correlation is generically expected because the shortest variability timescale is determined by the size of the blazar central engine. 

We have attempted to verify the existence of the "minimal timescale" -- "black hole mass" correlation,  using the data of Fermi/LAT. Our analysis shows that the LAT sensitivity is sufficient for detection of the variability timescales as short as the SMBH light crossing time and/or period of rotation around the last stable circular orbit for a number of bright flaring blazars, see Fig.~\ref{Results_types}. From this figure one could see that some $\sim 10$ blazars (out of 86 considered in the analysis)  exhibit variability on the "fundamental" SMBH timescales. At the same time, none of the sources is variable on the timescales shorter than the black hole light crossing time. 

From the same figure one could see that most of the sources are variable at least on the timescales within an order of magnitude from the period of rotation around the last stable orbit around Schwarzschild black hole. For most of the sources the minimal variability timescale measured by LAT is determined by the low statistics of the signal in LAT,  rather than by the physics of the \gr\ emission intrinsic to the source. This is clear from Fig.~\ref{Results_countrates}, in which a trend of the decrease of the minimal measured variability timescale with the increase of the peak flux is observed. Only three sources in our sample (3C 273, 4C +21.35 and PKS 1510-089) deviate from the formal sensitivity limit due to the finite \gr\ countrate in LAT, shown by the dashed line in Fig.~\ref{Results_countrates}. 

In these three sources the measured minimal variability timescale might be determined by the physics of the source, rather than by the LAT performance. In this case the measurement of the minimal variability timescale would provide an important information on the nature of the \gr\ emission. Indeed, the estimates of the masses of black holes in 4C +21.35, PKS 1510-089 and 3C 273 differ by more than an order of magnitude, but in all the three sources the minimal variability timescale derived from the LAT data is close to the light crossing time of the SMBH. This provides a support for the hypothesis of relation of the minimal variability timescale to the size of the blazar central engine. It is clear, however, that accumulation of larger sample of blazars, for which the intrinsic minimal variability timescale is measured, is needed before definitive conclusions on the validity of this hypothesis could be drawn. 

One possible caveat of the measurement of the minimal variability timescale in 4C +21.35, PKS 1510-089 and 3C 273 could be in the peculiarity of the observation mode of LAT. LAT images each particular direction on the sky for the time spans of $\sim 0.5$~hr once every  3.2 hours. Detection of variability on the timescales between 0.5 and 3.2 hours could, therefore, be affected by the irregularity of exposure by LAT. In the cases of PKS 1510-089 and 4C +21.35, the minimal detectable variability timescales are shorter than 0.5~hr so that, in principle, the variability could be measured within a single 0.5~hr exposure (with several time bins contained in this exposure). However, in the case of 3C 273, the minimal measurable variability timescale is close to 0.5~hr, so that the variability on this timescale might be missed if a particular flare of the source is not fully contained within a single 0.5~hr exposure. At the same time, 3C 273 produced a number of bright flares over the three year exposure with LAT, so that it is perhaps unlikely that half-an-hour scale variability has been missed in all the flares. Clarification of the issue of the instrumental 3.2~hr and 0.5~hr scales for the search of the minimal variability timescales will become possible with accumulation of more flaring sources with minimal detectable variability timescales in this range.   

If the observed minimal variability timescales are indeed related to the timescales of the SMBH in the blazar central engine, they provide a direct measurement of the size of the central engine, without an uncertainty related to the Doppler factor of the emitting plasma \citep{Neronov_times_scales}. Indeed, as the black hole and the observer are located in the same rest frame, the variability produced by the black hole will be seen by observer at the very same timescale, independently of what happens on the way with the medium, carrying the emission. Thus, the only difference between the intrinsic and observed timescales comes from the cosmological redshift of the source, which enlarges the observed timescales by $(1+z)$ with respect to the intrinsic ones. Thus, \gr\ observations could potentially provide independent estimates of the masses of SMBHs in the cores of AGN, providing a important tool in the context of the study of the formation and evolution of the SMBHs. Indeed, in our analysis only 86 sources with known estimates of the SMBH masses were included (see Table 1). However, LAT AGN catalog \citep{2FGL} includes a much larger number of variable sources, for which the measurements of the minimal variability timescales could be obtained, resulting in the estimates (or upper limits) on the masses of the SMBHs.  The \gr\ tool for the estimate of the black hole masses would be usable only after accumulation of larger statistics of the measurements of the intrinsic minimal variability timescales of \gr\ emission from blazars and verification of the possible correlation between the black hole mass and variability timescale shown in Fig.~\ref{Results_types}.

As it was mentioned before, current instruments, such as Fermi/LAT, allow only for the brightest objects to be studied at the sufficient level of details. The question whether the minimal variability timescale is indeed limited by the size of central black hole, thus, might need to await for the new generation of $\gamma$-ray instruments to become operational.

\bibliography{vovk}
\bibliographystyle{apj}

\end{document}

%% file: Table_of_sources.tex
\begin{longtable*}{llccccccc}
\hline \hline
   & Source name               &  Type  & RA       & Dec      & z      & $\log\left(\frac{\displaystyle \tau_{min}}{\displaystyle 1\mbox{ s}}\right)$ & $\log\left(\frac{\displaystyle M_{\rm BH}}{\displaystyle M_\odot}\right)$& Ref.\\
\hline
 1 &         1ES 0033+595     &  bzb   &    8.964 &   59.854 &  0.086 &  ---  & $7.25\pm0.70$ & 1 \\
 2 &         1ES 0120+340     &  bzb   &   20.665 &   34.420 &  0.272 &  ---  & $8.69\pm0.70$ & 1 \\
 3 & \textbf{1Jy 0138-097   } &  bzb   &   25.396 &   -9.481 &  1.034 &  6.03 & $9.63\pm0.70$ & 1 \\
 4 &         4C +15.05        &  agu   &   31.256 &   15.235 &  0.405 &  5.78 & $7.91\pm0.70$ & 3 \\
 5 & \textbf{PKS 0208-512   } &  agu   &   32.696 &  -51.035 &  0.999 &  5.61 & $8.78\pm0.70$ & 3 \\
 6 & \textbf{3C 66A         } &  bzb   &   35.662 &   43.036 &  0.444 &  5.10 & $8.30\pm0.70$ & 3 \\
 7 &         4C +28.07        &  bzq   &   39.472 &   28.778 &  1.207 &  ---  & $7.98\pm0.70$ & 3 \\
 8 &         AO 0235+164      &  bzb   &   39.675 &   16.624 &  0.940 &  5.94 & $7.98\pm0.70$ & 3 \\
 9 &         1H 0323+022      &  bzb   &   51.545 &    2.413 &  0.147 &  ---  & $8.36\pm0.70$ & 1 \\
10 &         PKS 0336-019     &  bzq   &   54.871 &   -1.744 &  0.852 &  6.02 & $7.21\pm0.70$ & 3 \\
11 &         1H 0414+009      &  bzb   &   64.209 &    1.092 &  0.287 &  ---  & $9.05\pm0.70$ & 1 \\
12 &         PKS 0420-01      &  bzq   &   65.807 &   -1.342 &  0.915 &  5.37 & $8.04\pm0.70$ & 3 \\
13 &         PKS 0440-00      &  bzq   &   70.686 &   -0.293 &  0.844 &  4.95 & $7.23\pm0.70$ & 3 \\
14 & \textbf{PKS 0454-234   } &  bzq   &   74.268 &  -23.427 &  1.003 &  4.83 & $8.05\pm0.70$ & 3 \\
15 &         PKS 0454-46      &  bzq   &   74.041 &  -46.218 &  0.853 &  ---  & $8.05\pm0.70$ & 3 \\
16 &         4C -02.19        &  bzq   &   75.319 &   -1.931 &  2.286 &  5.73 & $8.50\pm0.70$ & 3 \\
17 & \textbf{PKS 0521-36    } &  agn   &   80.769 &  -36.469 &  0.055 &  4.57 & $8.68\pm0.12$ & 2 \\
18 &         PKS 0528+134     &  bzq   &   82.714 &   13.553 &  2.070 &  ---  & $7.66\pm0.70$ & 3 \\
19 &         1Jy 0537-286     &  bzq   &   84.846 &  -28.697 &  3.104 &  6.21 & $7.78\pm0.70$ & 3 \\
20 &         PKS 0537-441     &  bzb   &   84.706 &  -44.084 &  0.894 &  6.04 & $8.02\pm0.70$ & 3 \\
21 &         1ES 0647+250     &  bzb   &  102.699 &   25.096 &  0.203 &  ---  & $7.73\pm0.70$ & 1 \\
22 &         EXO 0706.1+5913  &  bzb   &  107.629 &   59.139 &  0.125 &  ---  & $8.67\pm0.70$ & 1 \\
23 &         S5 0716+714      &  bzb   &  110.476 &   71.350 &  0.300 &  4.80 & $7.74\pm0.70$ & 1 \\
24 &         PKS 0735+17      &  bzb   &  114.524 &   17.703 &  0.424 &  6.05 & $8.08\pm0.70$ & 1 \\
25 &         PKS 0754+100     &  bzb   &  119.288 &    9.963 &  0.266 &  ---  & $8.21\pm0.70$ & 1 \\
26 &         1ES 0806+524     &  bzb   &  122.460 &   52.308 &  0.137 &  ---  & $8.65\pm0.70$ & 1 \\
27 &         S4 0814+425      &  bzb   &  124.573 &   42.398 &  1.073 &  ---  & $8.01\pm0.70$ & 1 \\
28 &         PKS 0823+033     &  bzb   &  126.478 &    3.144 &  0.506 &  ---  & $8.55\pm0.70$ & 1 \\
29 &         B2 0827+24       &  bzq   &  127.646 &   24.122 &  0.940 &  ---  & $8.41\pm0.43$ & 4 \\
30 &         PKS 0829+046     &  bzb   &  127.987 &    4.486 &  0.174 &  6.06 & $8.52\pm0.70$ & 1 \\
31 &         4C +71.07        &  bzq   &  130.420 &   70.880 &  2.172 &  4.41 & $7.22\pm0.70$ & 3 \\
32 & \textbf{OJ 287         } &  bzb   &  133.713 &   20.098 &  0.306 &  5.11 & $8.50\pm0.70$ & 1 \\
33 &         S4 0917+44       &  bzq   &  140.236 &   44.697 &  2.189 &  5.08 & $7.88\pm0.70$ & 3 \\
34 &         4C +55.17        &  bzq   &  149.433 &   55.382 &  0.895 &  5.50 & $8.42\pm0.70$ & 3 \\
35 &         S4 0954+658      &  bzb   &  149.652 &   65.557 &  0.368 &  5.67 & $8.37\pm0.70$ & 1 \\
36 &         1H 1013+498      &  bzb   &  153.788 &   49.432 &  0.212 &  ---  & $8.94\pm0.70$ & 1 \\
37 &         1ES 1028+511     &  bzb   &  157.761 &   50.895 &  0.360 &  ---  & $8.70\pm0.70$ & 1 \\
38 &         MKN 421          &  bzb   &  166.120 &   38.213 &  0.030 &  ---  & $8.23\pm0.70$ & 1 \\
39 &         1ES 1106+244     &  bzb   &  167.349 &   24.240 &  0.460 &  ---  & $8.64\pm0.70$ & 1 \\
40 &         PKS 1127-145     &  bzq   &  172.580 &  -14.803 &  1.187 &  ---  & $7.75\pm0.70$ & 3 \\
41 &         MKN 180          &  bzb   &  174.178 &   70.164 &  0.045 &  ---  & $8.10\pm0.70$ & 1 \\
42 &         4C +29.45        &  bzq   &  179.878 &   29.247 &  0.724 &  5.59 & $8.56\pm0.21$ & 4 \\
43 &         ON 231           &  bzb   &  185.374 &   28.239 &  0.102 &  ---  & $8.01\pm0.70$ & 1 \\
44 & \textbf{4C +21.35      } &  bzq   &  186.226 &   21.380 &  0.432 &  3.64 & $8.18\pm0.70$ & 3 \\
45 & \textbf{3C 273         } &  bzq   &  187.277 &    2.042 &  0.158 &  4.70 & $9.38\pm0.17$ & 5 \\
46 &         PG 1246+586      &  bzb   &  192.063 &   58.350 &  0.847 &  ---  & $9.15\pm0.70$ & 1 \\
47 &         3C 279           &  bzq   &  194.042 &   -5.794 &  0.536 &  5.48 & $7.82\pm0.70$ & 3 \\
48 &         1Jy 1418+546     &  bzb   &  215.067 &   54.376 &  0.153 &  ---  & $8.74\pm0.70$ & 1 \\
49 &         PG 1424+240      &  bzb   &  216.760 &   23.795 &  0.160 &  ---  & $6.42\pm0.70$ & 1 \\
50 &         PG 1437+398      &  bzb   &  219.802 &   39.536 &  0.344 &  6.25 & $8.95\pm0.70$ & 1 \\
51 &         1ES 1440+122     &  bzb   &  220.688 &   11.996 &  0.162 &  ---  & $8.55\pm0.70$ & 1 \\
52 &         BZB J1501+2238   &  bzb   &  225.275 &   22.639 &  0.235 &  ---  & $8.55\pm0.70$ & 1 \\
53 & \textbf{PKS 1510-089   } &  bzq   &  228.207 &   -9.103 &  0.360 &  3.84 & $8.31\pm0.27$ & 4 \\
54 &         1H 1515+660      &  bzb   &  229.517 &   65.437 &  0.702 &  ---  & $9.36\pm0.70$ & 1 \\
55 &         AP Lib           &  bzb   &  229.435 &  -24.360 &  0.049 &  ---  & $8.64\pm0.14$ & 2 \\
56 &         RGB J1534+372    &  bzb   &  233.861 &   37.335 &  0.143 &  6.42 & $8.13\pm0.70$ & 1 \\
57 &         PG 1553+113      &  bzb   &  238.942 &   11.190 &  0.360 &  ---  & $7.25\pm0.70$ & 1 \\
58 &         PKS 1604+159     &  bzb   &  241.770 &   15.876 &  0.357 &  ---  & $8.25\pm0.70$ & 3 \\
59 &         4C +10.45        &  bzq   &  242.148 &   10.494 &  1.226 &  ---  & $8.07\pm0.70$ & 3 \\
60 &         OS 319           &  bzq   &  243.372 &   34.159 &  1.397 &  ---  & $8.22\pm0.70$ & 3 \\
61 &         4C +38.41        &  bzq   &  248.809 &   38.171 &  1.814 &  4.81 & $7.54\pm0.70$ & 3 \\
62 &         MKN 501          &  bzb   &  253.481 &   39.763 &  0.034 &  ---  & $8.72\pm0.70$ & 1 \\
63 & \textbf{NRAO 530       } &  bzq   &  263.279 &  -13.128 &  0.902 &  4.90 & $8.07\pm0.70$ & 3 \\
64 &         4C +51.37        &  bzq   &  265.060 &   52.201 &  1.375 &  5.70 & $7.97\pm0.70$ & 3 \\
65 &         S4 1738+476      &  bzq   &  265.089 &   47.645 &  0.316 &  ---  & $8.22\pm0.70$ & 1 \\
66 &         1ES 1741+196     &  bzb   &  266.044 &   19.579 &  0.083 &  ---  & $8.93\pm0.70$ & 1 \\
67 & \textbf{1Jy 1749+701   } &  bzb   &  267.217 &   70.111 &  0.770 &  4.68 & $9.90\pm0.70$ & 1 \\
68 & \textbf{OT 081         } &  bzb   &  267.876 &    9.640 &  0.322 &  5.47 & $8.64\pm0.70$ & 1 \\
69 & \textbf{S5 1803+784    } &  bzb   &  270.147 &   78.483 &  0.680 &  4.95 & $8.82\pm0.70$ & 1 \\
70 &         B2 1811+31       &  bzb   &  273.395 &   31.722 &  0.117 &  6.26 & $8.82\pm0.70$ & 1 \\
71 &         4C +56.27        &  bzb   &  276.001 &   56.838 &  0.663 &  6.60 & $9.10\pm0.70$ & 1 \\
72 & \textbf{PKS 1830-210   } &  bzq   &  278.413 &  -21.075 &  2.507 &  4.44 & $8.05\pm0.70$ & 3 \\
73 &         PKS 1933-400     &  bzq   &  294.320 &  -39.933 &  0.966 &  ---  & $8.06\pm0.70$ & 3 \\
74 &         1ES 1959+650     &  bzb   &  300.020 &   65.157 &  0.047 &  ---  & $8.14\pm0.70$ & 1 \\
75 &         1Jy 2005-489     &  bzb   &  302.375 &  -48.834 &  0.071 &  6.48 & $8.57\pm0.14$ & 6 \\
76 &         PKS 2052-47      &  bzq   &  314.068 &  -47.255 &  1.489 &  5.57 & $7.88\pm0.70$ & 3 \\
77 & \textbf{4C -02.81      } &  bzb   &  323.466 &   -1.906 &  1.285 &  5.88 & $9.96\pm0.70$ & 1 \\
78 &         S3 2141+17       &  bzq   &  325.879 &   17.720 &  0.211 &  5.45 & $8.14\pm0.34$ & 4 \\
79 &         PKS 2149+173     &  bzb   &  328.102 &   17.590 &  0.000 &  ---  & $7.78\pm0.70$ & 1 \\
80 &         PKS 2155-304     &  bzb   &  329.715 &  -30.219 &  0.117 &  6.69 & $7.60\pm0.70$ & 3 \\
81 & \textbf{BL Lac         } &  bzb   &  330.707 &   42.268 &  0.069 &  4.96 & $8.58\pm0.70$ & 1 \\
82 &         PKS 2201+04      &  bzb   &  331.164 &    4.705 &  0.027 &  ---  & $7.76\pm0.13$ & 2 \\
83 &         B3 2247+381      &  bzb   &  342.512 &   38.419 &  0.119 &  6.15 & $8.64\pm0.70$ & 1 \\
84 & \textbf{3C 454.3       } &  bzq   &  343.497 &   16.153 &  0.859 &  2.94 & $6.76\pm0.70$ & 3 \\
85 &         BZB J2322+3436   &  bzb   &  350.672 &   34.588 &  0.098 &  ---  & $6.76\pm0.70$ & 3 \\
86 &         1ES 2344+514     &  bzb   &  356.759 &   51.705 &  0.044 &  ---  & $8.57\pm0.70$ & 1 \\
\hline \hline
\caption{List of the sources used in this work. Bold marks sources found to be variable on the timescales comparable with the light-crossing time of the central black hole and/or the period of the last stable orbit. Sources classification is that of \cite{2FGL}. References: (1) \cite{BLLac170}, (2) \cite{FundPlane}, (3) \cite{Fan_BH_masses}, (4) \cite{XieCat}, (5) \cite{Sephane_3C273}, (6) \cite{Wagner_cat}.}
\label{SrcTable}
\end{longtable*}